# Novel Fine-Structure in the Low-Energy Excitation Spectrum of a High-$T_c$ Superconductor by Polarization Dependent Photoemission

R. Manzke, R. Müller, C. Janowitz, M. Schneider, A. Krapf, H. Dwelk

*Humboldt-Universität, Institut für Physik, Invalidenstr. 110, D-10115 Berlin, Germany*

Angle-resolved photoemission spectroscopy is performed on single crystals of the single-layer high-$T_c$ superconductor $Bi_2Sr_{2-x}La_xCuO_{6+\delta}$ at optimal doping (x=0.4) in order to study in great detail the Zhang-Rice (ZR) singlet band at the Fermi level. Besides the high crystal quality the advantages of a single-layer material are the absence of bilayer effects and the distinct reduction of thermal broadening. Due to the high energy and angle resolution and, most important, due to the controlled variation of the polarization vector of the synchrotron radiation the emission from the ZR singlet band reveals a distinct fine-structure. It consists of two maxima, the first showing only weak and the second at $E_F$ extremely strong polarization dependence. However, our observation has enormous consequences for line shape analyses and the determination of pseudo gaps by photoemission.

The progress achieved by angle-resolved photoemission over the last decade in order to solve step by step the characteristics of high-$T_c$ superconductivity of the cuprates is impressive. The investigations of the weakly dispersing Cu-O derived ZR singlet band and the hole-like Fermi surface, studied extensively already in the early 90th, and nowadays Fermi liquid or non-Fermi liquid behavior by line shape analyses, the anisotropy of the superconducting gap and the observation of pseudo gaps [1][2] may be mentioned as examples. The latter has been found in photoemission for underdoped material, i.e. in a region of the phase diagram where it was theoretically proposed by the RVB model [3].

Besides this progress one should be aware that nearly all of our experimental photoemission knowledge of the cuprates bases on one material, namely Bi-2212. On YBCO surfaces for instance the superconductivity is found to be suppressed and the existence of surface states is discussed controversially [1]. Although Bi-2212 is without doubt the prototype high-$T_c$ material, due to its extraordinary good crystal quality, it also has some disadvantages. Only two points shall be mentioned : (i) An open question are still the bilayer effects. Proposed are band splits in several critical regions of the Brillouin zone, but no experimental verification has been reported yet. Even though these states can not be resolved experimentally they may provide additional broadening. Only nowadays, photoemission work of single-layer superconductors is increasing [4]-[6]. (ii) The relatively high critical temperature of Bi-2212 turns out to be in some sense spectroscopically disadvantageous, as one has to work at about 100 K sample temperature in order to study the normal state. Instead of this, the single-layer material Bi-2201 of optimal doping with $T_c$=29K allows a much lower working temperature and thermal broadening effects ($\approx 4k_BT_c$) are distinctly reduced below 10 meV.

In this investigation we present a detailed low-temperature photoemission study of the ZR singlet band near $E_F$ of single-layer optimal doped $Bi_2Sr_{2-0.4}La_{0.4}CuO_{6+\delta}$ in the normal state at 35 K. Besides high energy and angle resolution the transition matrix element is systematically varied by applying clear polarization geometries of the linearly polarized synchrotron radiation with respect to the high symmetry directions of the single crystals. In the way the matrix element is varied here it helps distinctly to suppress or enhance certain spectral features beyond the limits of experimental resolution. This will result in novel insight in the fine-structure of the ZR singlet band.

The single-layer $Bi_2Sr_{2-x}La_xCuO_{6+\delta}$ single crystals were grown out of the stoichiometric melt [7]. Lanthanum free samples (x=0) are found to be strongly hole overdoped with an $T_c$ of about 7 K. In order to reach optimal $T_c$ part of the $Sr^{2+}$ must be replaced by $La^{3+}$ what reduces the hole concentration in the Cu-O plane. The dependence of $T_c$ on x is roughly parabolic. The optimum $T_c$ of 29 K and the sharpest transition $\Delta T$ of 2 K is found for a La content of x=0.4 controlled by measuring the ac susceptibility (details will be published elsewhere [8]). The samples are rectangular shaped with the long side along the crystallographic a-axis, as confirmed by Laue diffraction and *in situ* LEED, and have a typical size of (5x2) mm$^2$. The crystal structure is orthorhombic due to the about 5x1 superstructure along the crystallographic b-axis [8]. The photoemission measurements reported in this paper have been carried out at the synchrotron radiation centers HASYLAB in Hamburg and BESSY in Berlin. At HASYLAB we used the WESPHOA-III station, equipped with a spherical analyzer on a two-axes goniometer, at the 3m normal-incidence monochromator HONORMI. The overall energy resolution was 40 meV, the angle resolution 1°. At BESSY we used the AR-65 station, equipped also with a spherical analyzer on a two-axes goniometer, which resolution has been tested to reach 8.5 meV [9]. The crossed undulator/monochromator combination U2/FSGM allows a rotation of the electrical field vector **E** by 90° [10]. These are optimal conditions for polarization dependent measurements. Photon energy changes between both polarizations are below 2 meV what has been tested at the Fermi-Dirac distribution of a Au film in electrical contact with the sample. This serves also for the determination of the Fermi energy $E_F$. The overall energy resolution was 30 meV, the angle resolution 1°. All spectra shown in Fig. 1 and 2 are absolutely normalized with respect to the photon flux. Thus, the intensities are directly comparable.

In Fig. 1 we show polarization dependent spectra series about half the distance between Γ (0,0) and X ($\pi$, $\underline{\pi}$) (left columns) and Γ and Y ($\pi$, $\underline{\pi}$) (right columns) of the Brillouin zone [11]. The **k** values are selected to be nearby the Fermi level crossing of the ZR singlet band. A very similar polarization dependent series has been given for the double-layer material Bi-2212 by Dessau et al. [12], discussed in detail in [1]. Polarization dependent series of Bi-2201 around the M-point ($\pi$, 0) varying also the photon energy can be found in Ref. [6].

Similar to the n=2 data discussed by Shen and Dessau [1] the polarization dependent spectra of Fig 1 can be principally analyzed by regarding the influence of the dipole part of the transition matrix element $M_{if}(\mathbf{k}) = (e/mc)<\Psi_f(\mathbf{k})|\mathbf{A}\cdot\mathbf{p}|\Psi_i(\mathbf{k})>$, with the vector potential of the light **A**. **A** is parallel to the electrical field vector **E**, the momentum of the photoelectron is denoted by **p**, and the initial and final states are $\Psi_i(\mathbf{k})$ and $\Psi_f(\mathbf{k})$, resp.. The polarization dependence of the photoemission intensity,

$$I(\mathbf{k},\omega) = \Sigma_i |M_{if}(\mathbf{k})|^2 f(\omega) A(\mathbf{k},\omega) \qquad (1),$$

with the Fermi function $f(\omega)$ and the initial state spectral function $A(\mathbf{k},\omega)$, is determined by the observable quantity $|M_{if}(\mathbf{k})|^2$ which must be invariant for the polarization geometry applied in the photoemission experiment. Such an approach is commonly used to understand

(or determine) the symmetry of the initial states contributing to the photoemission intensity. In the particular case of the Bi-2201 spectra series of Fig. 1 the analysis reveals similar results as for Bi-2212 [1], meaning that the polarization dependence is qualitatively consistent with the assumption of an initial state of $d_{x^2-y^2}$ symmetry. Note that the consistence with the data also holds for $\Psi_i$ of $p_{x,y}$ symmetry. This is directly clear when one regards the overlap (bonds) of the hybridized O $p_{x,y}$ and Cu $d_{x^2-y^2}$ states in the Cu-O planes.

Besides these similarities with the Bi-2212 data [1] the Bi-2201 spectra series reveal distinct differences: At first, for Bi-2212 the spectra along $\Gamma X$ were always more intense than along $\Gamma Y$. For Bi-2201 we observe the opposite behaviour (Fig. 1). At second, regarding the spectra of the two polarization geometries of both momentum directions in more detail, one realizes that the effect of polarization is not solely an intensity variation. The intensity increase is combined, on the one hand, with a distinct sharpening of the spectral line and, on the other hand, with an energy shift of the emission maximum. For certain polarizations, e.g. left column for $\Gamma X$ and right column for $\Gamma Y$ of Fig. 1, the spectral weight at $E_F$ is distinctly suppressed. Regarding in these cases the point of half height of the spectral onset, it is clearly shifted away from $E_F$ to higher binding energies. As this method is commonly used to derive an energy gap (defined by the energy difference between the half height of the spectral onset and $E_F$) from photoemission spectra, it becomes obvious that this must result into misleading interpretations if the polarization conditions are not exactly clear. In addition, we would like to point attention, for instance, on the two spectra taken with an emission angle of 16° along $\Gamma Y$ (column 3 and 4 of Fig. 1). By changing the polarization direction the spectral weight at $E_F$ can be switched on and off.

In order to improve the clean polarization conditions further we studied again a spectra series along ΓX and supplementary along ΓM at the undulator beamline at BESSY where the direction of **E** can be changed directly by the undulator (see the experimental part above), i.e. without moving the sample or the analyser for the two polarization series shown in Fig. 2. Only **E** is varied by 90°. The dramatic polarization effect due to the 90° tilt is obvious. Although, the intensity variations for the ΓX series again could be principally explained by matrix element changes discussed above, the energy shifts between the two polarization dependent spectral contributions surely can not. In addition, for the ΓM series shown in Fig. 2 already the intensity variations violate the selection rules of $M_{i\,f}(\mathbf{k})$ which is identical in both geometries. As the spectra are absolutely normalized, the almost vanishing polarization dependence of the emission maximum (or spectral line) at higher $E_B$ is a true effect. This is additionally demonstrated by the dashed contributions. The polarization strongly acts on the second spectral line at $E_F$, almost switching it on and off, and the variation of this line is given by the difference spectra shown in Fig. 2. It should be mentioned that such a polarization dependence found here is much stronger than normally observed in solids, e.g. the dangling bond states on semiconductor surfaces [13].

Regarding the polarization dependent series of Fig. 2 and, in particular, the difference spectra this reveals distinct novel fine-structure of the ZR singlet band around M(π,0) as well as about half ΓX(π,π). According to equ. 1 it must consist of two spectral functions, one near $E_F$ (assigned S) and a second at some higher $E_B$ (assigned H) revealing strong and weak polarization dependence, resp.. The spectral function S has minimum $E_B$ of 10.5 meV at M and its width of 35 meV is limited by experimental resolution (including thermal broadening).

The origin of the double-peak structure at $E_F$ revealing the peculiar polarization properties can not be due to bilayer effects, because of the single-layer material, and is less probably for effects of the electronic band structure. According to band structure calculations [14] a second Bi-O derived band or Fermi surface is proposed around the M-point, serving for the doping of the Cu-O planes. For Bi-2201 significant differences between the tetragonal and orthorhombic crystal structures have been found in the calculations [15] around the M point (our notation [11]). Although an additional Bi-O derived band may account for a double-peak structure in the spectra at M, it should be absent along ΓX and ΓY, where we observed also two peaks, although less pronounced. In addition, one would not expect such strong polarization dependencies and for Bi-O initial states one would derive at similar conclusions for the symmetry selection rules of the transition matrix element like for the Cu-O initial states discussed above. But there is one asymmetry present in all Bi-cuprates, the approximately 5x1 superstructure along the crystallographic b-axis. It is up to now not clear whether such an asymmetry may account for the observed effects. Another scenario might be to think of the double-peaked spectral function as a hint of the separation of the spin and charge degree of freedom, similar like for a 1D Luttinger liquid [3]. In this case emission maximum H would correspond to a diffuse charge peak, the extremely sharp maximum S at $E_F$ to the spin (or spinon) peak. For the description of the marked polarization dependence of the latter the extension of equ. 1 by relativistic dipole selection rules will be probably necessary. In certain geometries it has been shown that relativistic selection rules must be applied even for nonmagnetic materials like Copper [16].

In summary, the novel spectral features and, especially, the strong dependence of the spectral function S at $E_F$ on the electrical field vector **E** observed and discussed in the present polarization dependent ARPES work is not understandable within the conventional frame work of photoemission and possibly points to an up to now disregarded effect of high-$T_c$ superconductivity in the electronic structure. This asks for further theoretical explanations. Nevertheless, our work demonstrates in several aspects the importance of the polarization geometry. Disregarding the polarization will by sure result in misleading interpretations, e.g. in the determination of pseudo gaps and the analysis of line-shapes of the spectral function at $E_F$.

We gratefully acknowledge assistance by the staff of HASYLAB, namely Dr. P. Gürtler, and of BESSY. We also thank the staff of the WESPHOA spectrometer of the University of Kiel, namely K. Roßnagel. For the crystal growth we thank D. Kaiser. This work was supported by the German Ministry of Science and Technology, project no. BMBF 05 SB8 KH1 0.

**Figure Captions**

FIG. 1.  Series of high-resolution photoemission spectra of $Bi_2Sr_{2-0.4}La_{0.4}CuO_{6+\delta}$ in the normal state at 35K. Shown are the $\Gamma X$ (left) and $\Gamma Y$ direction (right) of the Brillouin zone around Fermi energy crossing for 18 eV photon energy. In the insets the direction of the polarization of the light and the measuring directions (by the arrows) are given.

FIG. 2.  $\Gamma X$ (a) and $\Gamma M$ (b) series of high-resolution photoemission spectra for two different polarization geometries (shown in the insets) of $Bi_2Sr_{2-0.4}La_{0.4}CuO_{6+\delta}$ in the normal state at 35K. The photon energy was 34 eV. Note that in this experiment the sample is fixed and the direction of the vector potential of the synchrotron light is switched by 90 °. The dashed contributions in the $\mathbf{E}\perp\Gamma X$ series are due to the intensities of the spectra with $\mathbf{E}\|\Gamma X$. The third column of (b) shows the difference spectra $I(\mathbf{E}\perp\Gamma X)$ minus $I(\mathbf{E}\|\Gamma X)$.

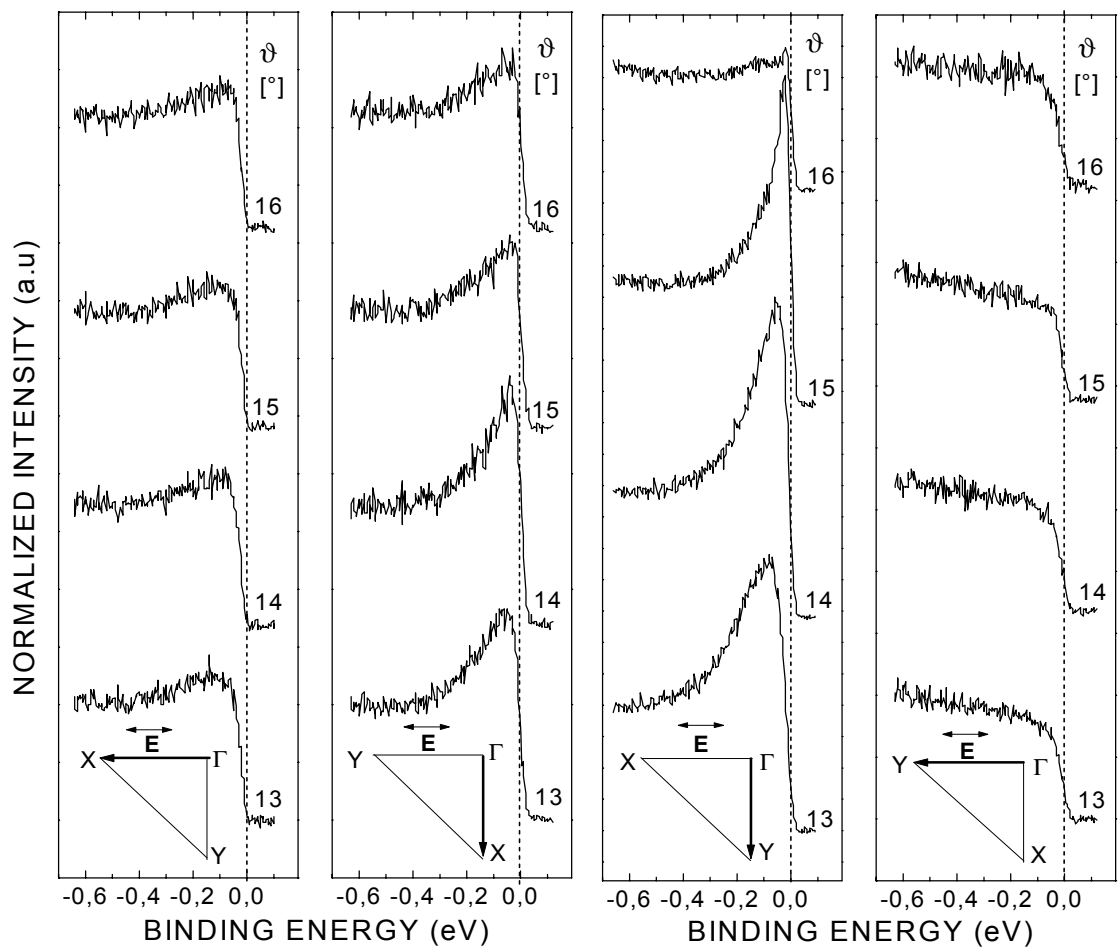

FIG. 1.
11

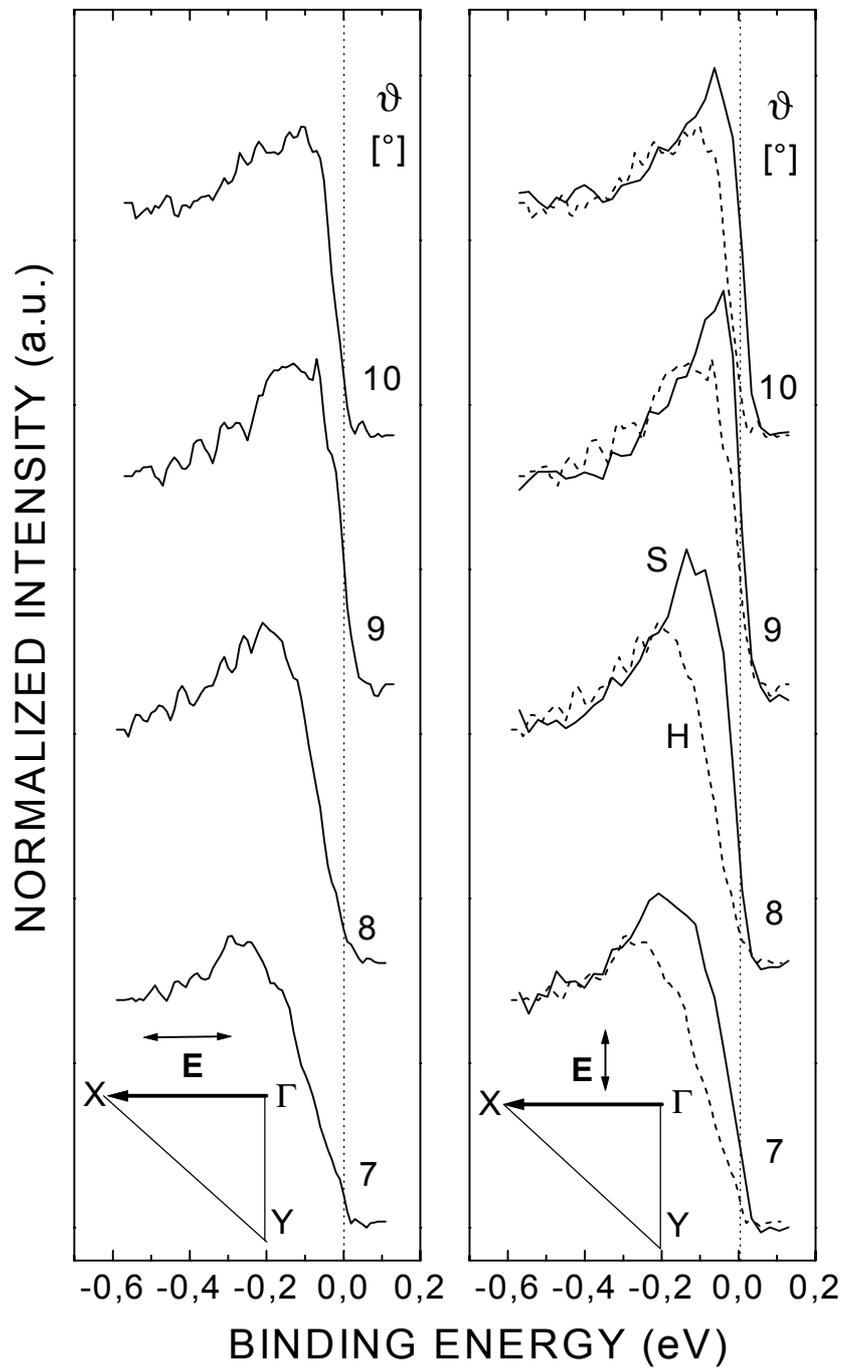

FIG. 2a

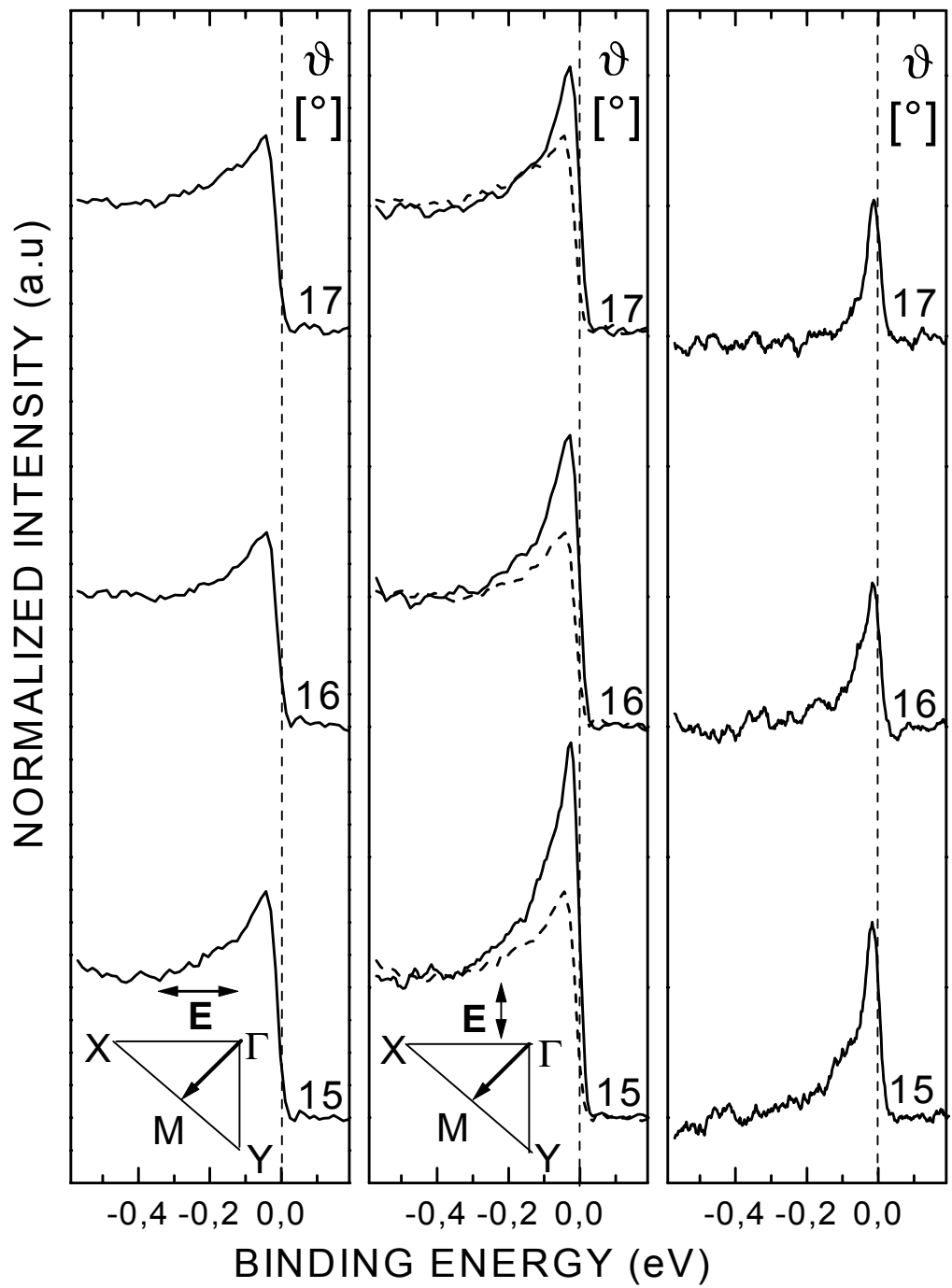

FIG.2b